\documentclass[aps,twocolumn,showpacs,pra,superscriptaddress,floatfix,longbibliography]{revtex4-1}

\usepackage{natbib}
\usepackage[breaklinks]{hyperref}
\usepackage{xstring}
\hypersetup{colorlinks=true}

\usepackage{amsmath,amssymb,amsthm,bm,graphicx,xcolor,amsfonts}
\newtheorem{thm}{Theorem}

\newtheorem{rem}{Remark}
\newtheorem{exmp}{Example}

\usepackage{color}
\usepackage{tabularx}
\usepackage{epsfig}

\usepackage{amssymb}
\usepackage{graphicx}
\usepackage{wasysym}
\usepackage{dcolumn}
\usepackage{epstopdf}
\usepackage{subfigure}
\usepackage{psfrag}

\newcommand{\vecform}{\bm}              

\newcommand{\R}{{\mathbb{R}}}
\newcommand{\la}{{\lambda}}
\newcommand{\up}{{\uparrow}}
\newcommand{\dn}{{\downarrow}}
\newcommand{\bra}[1]{\mbox{$\langle #1 |$}}
\newcommand{\bd}[1]{\boldsymbol{ #1 }}
\newcommand{\ket}[1]{\mbox{$| #1 \rangle$}}

\renewcommand{\H}{\mathcal{H}}

\newcommand{\rr}{\vecform{r}}

\begin{document}

\title{Natural Extension of Hartree-Fock through extremal $1$-fermion information: Overview and application to the lithium atom}

\author{Carlos L. Benavides-Riveros}
\email{carlos.benavides-riveros@physik.uni-halle.de}
\affiliation{Institut f\"ur Physik, Martin-Luther-Universit\"at Halle-Wittenberg, 06120 Halle, Germany}

\author{Christian Schilling}
\email{christian.schilling@physics.ox.ac.uk}
\affiliation{Clarendon Laboratory, University of Oxford, Parks Road, Oxford OX1 3PU, United Kingdom}

\date{\today}

\begin{abstract}
Fermionic natural occupation numbers do not only obey Pauli's
exclusion principle but are even stronger restricted by so-called
generalized Pauli constraints. Whenever given natural occupation
numbers lie on the boundary of the allowed region the corresponding
$N$-fermion quantum state has a significantly simpler structure.
We recall the recently proposed natural extension of the Hartree-Fock
ansatz based on this structural simplification. This variational ansatz is
tested for the lithium atom. Intriguingly, the underlying mathematical
structure yields universal geometrical bounds on the correlation
energy reconstructed by this ansatz.
\end{abstract}

\pacs{31.15.V-, 03.67.-a, 05.30.Fk}


\maketitle

\begin{centering}
\textit{To Michael Springborg, on his 60th anniversary.}
\end{centering}

\section{Introduction}\label{sec:intro}

In January 1925, Wolfgang Pauli announced the famous principle
which takes his name \cite{Pauli1925}, stating that no two identical
fermions can occupy the same quantum state at the same time.
It is difficult to underestimate its importance in physics and chemistry.
For instance, the Pauli exclusion principle (PEP) explains
the classification of atoms in the periodic table, the electronic structure
of atoms and molecules and is essential for the stability of matter.

Originally, PEP was introduced as a phenomenological rule
to explain some known spectroscopic anomalies \cite{Micaela}.
Yet, already in 1926, Dirac \cite{Dirac1926} and Heisenberg \cite{Heis1926}
provided a justification for it. They identified PEP as a consequence
of the antisymmetry of $N$-fermion wavefunctions under particle
exchange.

In quantum chemistry and quantum physics, due to the exponential
scaling of Hilbert spaces as the system size increases, one is keen
to avoid the use of wave functions. Indeed, since the physical
systems considered there have only $1$- and $2$-particle
interactions it is quite promising to restrict oneself to the corresponding
$n$-body reduced density matrices ($n$-RDM), with $n=1,2$.
In general, the $n$-RDM is defined as the contractions of an $N$-body
density matrix $\rho_N$ (see, e.g., Ref.~\cite{Davidson76}),
 \begin{equation}\label{nrdo}
\rho_n\equiv \binom{N}{n}\,\mbox{tr}_{N-n}[\rho_N]\,.
\end{equation}
Here, $\rho_N$ might be pure, $\rho_N\equiv \ket{\Psi}\bra{\Psi}$ or
just an \textit{ensemble} state. Due to the exchange symmetry the $n$-RDM
(\ref{nrdo}) is independent of the choice of the $N-n$ particles which are
traced out.

In this context, a more modern and more general version of the PEP can be formulated: The natural
occupation numbers (NON), the eigenvalues of the $1$-RDM,
 can be no larger than $1$,
\begin{equation}\label{eq:PEP}
0\leq \lambda_i\leq 1\,.
\end{equation}
This upper bound for the spin-orbital occupancies allows no more
than one electron in each quantum state. This elementary condition,
concluded by Coleman in 1963 \cite{Coleman}, is necessary and
sufficient for a $1$-RDM $\rho_1$ to be compatible
with an \textit{ensemble} $N$-fermion state $\rho_N$, provided
the trace condition holds, $\mbox{tr}_1[\rho_1]=N$. In other words,
$\rho_1$ is the contraction of a fermionic $\rho_N$ if and only if all eigenvalues of
$\rho_1$ obey Eq.~(\ref{eq:PEP}).

The physical relevance of PEP has a twofold origin: For many fermionic
quantum systems most of their NON do (approximately) saturate PEP
(\ref{eq:PEP}), i.e.~one observes either $\lambda_i\approx 1$ or
$\lambda_i\approx 0$. This quasipinning of most $\lambda_i$ to the
minimal or maximal values leads to a significant simplification of the
theoretical description of the physical system: Fermions in lower lying
energy shells are frozen and higher shells can be neglected. Moreover,
the behavior of the system can then be described by strongly reduced
active spaces \cite{CASSCF1,CASSCF2,CASSCF3,CASSCF4}. A prime
example for such a reduction of degrees of freedom is the celebrated
Hartree-Fock approximation. It is defined by the assumption that each
NON is saturating the PEP, i.e.~\begin{equation}\label{eq:HF}
\vec{\lambda}\equiv(\lambda_i)=\vec{\lambda}_{\rm HF} \equiv (\underbrace{1,\ldots,1}_N,0,\ldots)\,.
\end{equation}
Indeed, it can easily be shown that such NON imply that the corresponding
$N$-fermion quantum state can be written as a single Slater determinant,
$\ket{\Psi}=\ket{1,2,\ldots,N}$ with some appropriate $1$-particle states $\{\ket{i}\}_{i=1}^N$.

\begin{figure}[t]
\centering
\includegraphics[width=9cm]{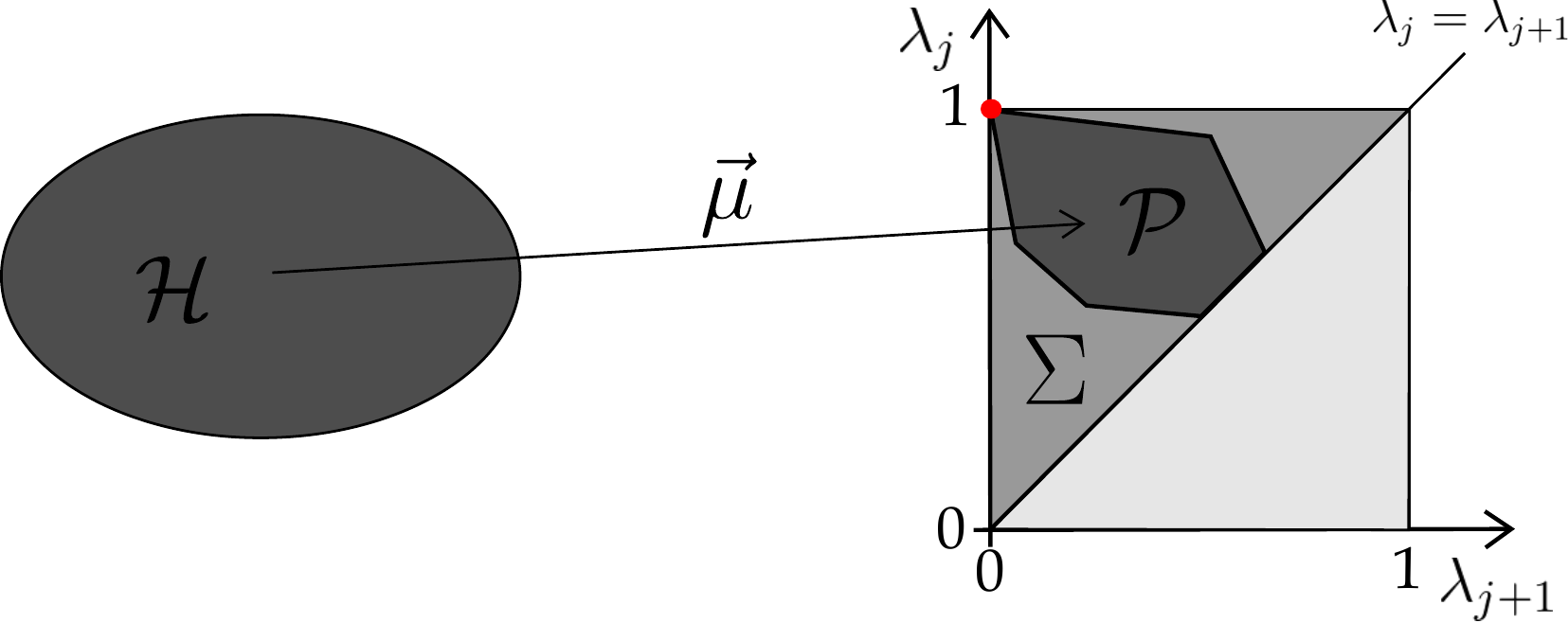}
\caption{Schematic illustration of the polytope $\mathcal{P}$
of vectors $\vec{\lambda}=(\lambda_i)$ of decreasingly ordered natural occupation numbers defined by the family of generalized Pauli constraints. Only those $\vec{\lambda}$ lying in $\mathcal{P}$ can arise from pure $N$-fermion quantum states \mbox{$\ket{\Psi}\in \mathcal{H}$}. Hartree-Fock point (\ref{eq:HF}) is shown as red dot.}
\label{fig:polytope1}
\end{figure}

Given that the PEP, beyond its fundamental nature, implies such strong structural implications for the wave function, researchers had been seeking  generalizations of it. This has led to the $1$-body $N$-presentability problem, the problem of describing all $1$-RDM which are compatible with pure $N$-fermion quantum states. In the 1970s, it was observed that further linear constraints on the NON for very specific settings of $N=3,4$ fermions emerge as a consequence of the global antisymmetry \cite{Borl1972}. Only recently, it was conclusively shown that the fermionic exchange symmetry implies \emph{in general} so-called \textit{generalized Pauli constraints} (GPC) \cite{Kly2,Kly3,Altun}, taking the form
\begin{equation}\label{eq:gpc}
D_j(\vec{\lambda}) \equiv \kappa_j^{(0)}+\sum_{i=1}^d\kappa_j^{(i)} \lambda_i\geq 0\,.
\end{equation}
Here $\kappa_j^{(i)} \in \mathbb{Z}$, $j=1,2,\ldots,\nu_{N,d}<\infty$, $d$ is the dimension of the underlying $1$-particle Hilbert space and the NON are decreasingly ordered. From a geometrical viewpoint, for each fixed pair $(N,d)$, the family of GPC, together with the normalization and the ordering constraints $\lambda_1\geq\ldots\geq \lambda_d\geq 0$, form a polytope $\mathcal{P}_{N,d}$ of allowed vectors $\vec{\lambda}\equiv (\lambda_i)_{i=1}^d$. This is also illustrated in Figure \ref{fig:polytope1}.
There the map $\vec{\mu}$ maps every $N$-fermion quantum state to its vector $\vec{\lambda}$ of decreasingly ordered NON. Here and in the following indices `($N$,$d$)' are skipped. Clearly, only those $\vec{\lambda}$ inside the `Pauli simplex' $\Sigma$, defined by $1\geq \lambda_1\geq \ldots\geq \lambda_d\geq 0$, can be reached. Yet, since $\mathcal{P} \subsetneq \Sigma$, the GPC are more restrictive than PEP. In Ref.~\cite{CSQ} the redundancy of the PEP constraints given all GPC was completely explored and quantified.

As an example, we consider the Borland-Dennis setting $\wedge^3[\H_1^{(6)}]$.
The GPC for the decreasingly ordered NON $(\lambda_i)_{i=1}^6$ read \cite{Borl1972,Rus2}
\begin{eqnarray}
&&\lambda_1 + \la_6 = \la_2 + \la_5 = \la_3 + \la_4 =1 \label{eq:gpc36a}\\
&& 2-(\la_1 + \la_2 + \la_4) \geq 0. \label{eq:gpc36b}
\end{eqnarray}
GPC (\ref{eq:gpc36b}) clearly exceeds PEP which states $\lambda_1+\lambda_2\leq 2$. That some GPC take the form of equalities rather than proper inequalities is quite unique. Besides the Borland-Dennis setting, this happens only for the settings $(N,d)$ with $N\leq 2$ or $d-N\leq 2$, i.e.~for the case of at most two fermions or at most two fermionic holes.

Particular relevance of GPC is given whenever the NON $\lambda_i$ of some system are saturating a GPC, i.e.~$\vec{\lambda}$ lies on the boundary of the polytope. This so-called \emph{pinning} effect and its consequences are discussed in detail in the succeeding section.

Finally, it should also be stressed that the solution of the \textit{pure} $N$-representability
problem for the $1$-RDM was part of a more general effort in quantum information theory \cite{Higushi,Brav,Kly4,MC,Daft}, addressing the so-called quantum marginal problem (for a rudimentary overview see \cite{CSQMath12}). Moreover, there is no algorithm known yet which allows one to determine the family of GPC for the setting $(N,d)$ efficiently in $d$.

The paper is arranged as follows. In Section \ref{sec:relevanceGPC} we briefly explain the potential physical relevance of GPC. The remarkable structural implications for $\ket{\Psi}$ in case of pinning are described in Section \ref{sec:ci}. In Section \ref{sec:mscsf} the multiconfigurational self-consistent field ansatz worked out in Ref.~\cite{CSHF} based on this pinning structure is recalled and tested for the lithium atom in Section \ref{sec:Li}. In Section \ref{sec:QPCorrE} the numerical quality of such ansatzes are shown to be strongly related to geometric distances in the $1$-particle picture (polytope).

\section{Potential physical relevance of generalized Pauli constraints and pinning/quasipinning-effect}\label{sec:relevanceGPC}

Direct relevance of GPC was suggested in \cite{Kly1,Kly5}: At least for some systems the ground state minimization process of the energy expectation value $\bra{\Psi_N}\hat{H}\ket{\Psi_N}$, from the viewpoint of the $1$-particle picture, may get stuck on the polytope boundary $\partial \mathcal{P}$ since any further minimization would violate some GPC. This \emph{pinning} effect is relevant because it can restrict the dynamics of the corresponding system since its NON $\vec{\lambda}$ can never leave the polytope. This is a generalization of the fact that, e.g., electrons in atoms cannot decay to lower lying energy shells since those are already occupied.

On the other hand, pinning as an effect in the $1$-particle picture allows one to reconstruct the structure of the corresponding $N$-fermion quantum state. In addition this structure is significantly simplified \cite{CSQMath12,Kly1,CSthesis,CSQuasipinning}. 

In a first analytic study \cite{CS2013}, however, strong evidence was provided that $\vec{\lambda}$ for ground states of interacting fermions lies extremely close to, but not exactly on the polytope boundary. For this conceptually different \emph{quasipinning} the same important implications of pinning hold approximately. Whether quasipinning is generic or appears only for specific systems is part of an ongoing debate \cite{CSQ,Kly1,CSthesis,CSQuasipinning,BenavLiQuasi,Mazz14,BenavQuasi2,RDMFT,Alex,CS2015Hubbard,RDMFT2,BenavThesis,RefNew2,RefNew1}.

Although pinning is expected to be quite idealized and unrealistic it was observed in Refs.~\cite{CSthesis,BenavQuasi2,CS2015Hubbard}
that it can occur as a consequence of $1$-particle symmetries. Also converse, the occurrence of pinning reveals a symmetry of the corresponding $N$-fermion quantum state $\ket{\Psi}$ \cite{CS2015Hubbard}.

As a caveat concerning further investigations of possible (quasi)pinning it should be noted that (quasi)pinning by GPC follows already from (quasi)pinning by PEP constraints: Since $\mathcal{P}\subset \Sigma$ (recall Figure \ref{fig:polytope1}) (approximate) saturation of some PEP constraints always implies (approximate) saturation of some GPC \footnote{Recall, that the vector $\vec{\lambda}$ of NON has always to lie inside the polytope $\mathcal{P}$.}. Hence, the question is whether some GPC are stronger saturated than one could expect from possible (approximate) saturation of PEP constraints. For this purpose, to quantify such relevance of GPC \emph{beyond} the well-known relevance of PEP a corresponding geometric quasipinning measure was constructed in Ref.~\cite{CSQ}, the so-called $Q$-parameter.

\section{Structural implications for $\ket{\Psi}$ from extremal natural occupation numbers}\label{sec:ci}
In this section we recall the implications of pinning for the structure of the corresponding $N$-fermion quantum state (for more details see Ref.~\cite{CSQuasipinning}). For this we exploit a self-consistent expansion for $N$-fermion quantum states \cite{CSQuasipinning}. For every $\ket{\Psi} \in \wedge^N[\mathcal{H}_1^{(d)}]$ we determine the corresponding $1$-RDM $\rho_1$ which can be diagonalized,
\begin{equation}\label{1rdo}
\rho_1\equiv N\,\mbox{tr}_{N-1}[\ket{\Psi}\langle\Psi|]\equiv\sum_{i=1}^d\lambda_i\ket{i}\langle i|.
\end{equation}
For given $\ket{\Psi}$ its \emph{natural spin orbitals}, the eigenvectors
$\ket{i}$ of $\rho_1$,  define an orthonormal basis $\mathcal{B}_1$
for the $1$-particle Hilbert space $\H_1^{(d)}$. Consequently, $\ket{\Psi}$ also induces a basis $\mathcal{B}_N$ for the
$N$-fermion Hilbert space $\wedge^N[\H_1^{(d)}]$ given by all $N$-fermion Slater determinants,
\begin{equation}
\ket{i_1,\ldots,i_N} \equiv \mathcal{A}_N \,\ket{i_1}\otimes\ldots\otimes\ket{i_N}\,,
\end{equation}
$1\leq i_1<\ldots<i_N\leq d$, $\ket{i_k} \in \mathcal{B}_1$ and $\mathcal{A}_N$ is the antisymmetrizing operator for $N$ particles.
$\ket{\Psi}$ can then be expanded with respect to $\mathcal{B}_N$,
\begin{equation}\label{eq:selfconsistent}
\ket{\Psi}=\sum_{\bd{i}}c_{\bd{i}}\,\ket{\bd{i}}\,,
\end{equation}
where $\bd{i}\equiv (i_1,\ldots,i_N)$.
\begin{rem}\label{rem:selfconsistent}
The coefficients $c_{\bd{i}}$ in Eq.~(\ref{eq:selfconsistent}) are not free,
but need to fulfill strong self-consistency conditions to ensure that
\begin{itemize}
\item the $1$-RDM is diagonal with respect to the natural spin orbitals $\ket{i}$
\item the NON are correctly ordered, $\lambda_i\geq \lambda_{i+1}$ for all $i$
\end{itemize}
\end{rem}

To motivate the strong structural implications for $\ket{\Psi}$ whenever
its NON are saturating a GPC, consider NON all saturating the PEP, i.e.~$\vec{\lambda}$
coincides with the Hartree-Fock point (\ref{eq:HF}). The corresponding $N$-fermion
quantum state $\ket{\Psi}$ is then given by a single Slater determinant $\ket{\Psi}=\ket{1,\ldots,N}$.
This strong structural implication of pinning to the Hartree-Fock point generalizes to pinning of NON to \emph{arbitrary}
points on the polytope boundary (recall also Figure \ref{fig:polytope1}). To explain this, assume that a GPC (\ref{eq:gpc})
is pinned,
\begin{align}
D_{j}(\vec{\lambda}) =0\,.
\label{eq:pinned}
\end{align}
A mathematical theorem then shows that any compatible $N$-fermion state $\ket{\Psi}$ (i.e.~with NON $\vec{\lambda}$) belongs
to the \mbox{$0$-eigenspace} (a proof is presented in \cite{Alex}) of the associated operator
\begin{align}\label{eq:operator}
\hat{D}_{j} =  \kappa^{(0)}_j + \kappa^{(1)}_j \hat{n}_1
+ \ldots + \kappa^{(d)}_j \hat{n}_d\,.
\end{align}
Here, $\hat{n}_i$ denotes the particle number operator for the natural spin orbital $\ket{i}$ of $\ket{\Psi}$ and the implicit dependence of $\hat{n}_i$ and $\hat{D}_j$ on given $\ket{\Psi}$ is suppressed.
This condition,
\begin{equation}
D_j(\vec{\lambda})=0 \quad \Rightarrow \quad \hat{D}_j\ket{\Psi}=0
\end{equation}
gives rise to a \textit{selection rule} for the Slater determinants
in the self-consisted expansion (\ref{eq:selfconsistent}) \cite{Kly1},
\begin{equation}\label{eq:selection}
\hat{D}_j\ket{\bd{i}} \neq 0\quad\Rightarrow\quad c_{\bd{i}}=0\,.
\end{equation}

By denoting the set of all configurations fulfilling such a
selection rule for a GPC $D$ by $\mathcal{I}_D$, pinning by $D$ simplifies expansion (\ref{eq:selfconsistent}) to
\begin{equation}\label{eq:MCSCF}
\ket{\Psi} = \sum_{\bd{i}\in \mathcal{I}_D}c_{\bd{i}}\ket{\bd{i}}\,.
\end{equation}

To illustrate these structural simplifications we recall an elementary example from Ref.~\cite{CSQuasipinning}.
\begin{exmp}\label{example1}
For the Borland-Dennis setting $\wedge^3[\H_1^{(6)}]$ the first three
GPC (\ref{eq:gpc36a}) take the form of equalities (independent of $\vec{\lambda}$)
which implies universal structural simplifications for any $\ket{\Psi} \in \wedge^3[\H_1^{(6)}]$.
They read
\begin{align}
(\hat{1} - \hat{n}_1 - \hat{n}_6) \ket{\Psi} &= 0,  \nonumber \\
(\hat{1}  - \hat{n}_2 - \hat{n}_5) \ket{\Psi} &= 0, \nonumber \\
(\hat{1} - \hat{n}_3 - \hat{n}_4) \ket{\Psi} &= 0\,,
\label{selBD}
\end{align}
where $\hat{1}$ denotes the identity operator.
Hence, only the configurations $\{i_1,i_2,i_3\}$ (with
$i_1 \in \{1,6\}$, $i_2 \in \{2,5\}$ and $i_3 \in \{3,4\}$)
can contribute to $\ket{\Psi}$ in the expansion \eqref{eq:selfconsistent}.
This results in the following eight possible configurations: $\ket{1,2,3}$, $\ket{1,2,4}$, $\ket{1,3,5}$, $\ket{1,4,5}$,
$\ket{2,3,6}$, $\ket{2,4,6}$, $\ket{3,5,6}$, $\ket{4,5,6}$.

If in addition GPC (\ref{eq:gpc36b}) is saturated $\ket{\Psi}$ also needs to meet
\begin{equation}
(2\cdot\hat{1}  - \hat{n}_1 - \hat{n}_2- \hat{n}_4) \ket{\Psi} = 0\,.
\end{equation}
The set of possible Slater determinants reduces to just three and the most generic quantum state with NON pinned by GPC (\ref{eq:gpc36b}) takes the form
\begin{align}
\label{eq:BDstate}
\ket{\Psi}= \alpha \ket{1,2,3}+ \beta  \ket{1,4,5} +\gamma \ket{2,4,6}.
\end{align}
Here, $|\beta| \geq |\gamma|$ and $|\alpha|^2 \geq |\beta|^2 + |\gamma|^2$ to meet the second self-consistency
condition in Remark \ref{rem:selfconsistent}.
\end{exmp}

\section{Variational ansatz and very first results}\label{sec:mscsf}

As suggested in Ref.~\cite{CS2013} and worked out in Ref.~\cite{CSHF}
the structural simplifications (\ref{eq:MCSCF}) can be used as a variational
ansatz. There, the energy expectation value
\begin{equation}\label{eq:MCSCFenergy}
E[\{c_{\bd{i}}\}_{\bd{i}\in \mathcal{I}_D},\{\ket{i}\}] \equiv \bra{\Psi}\hat{H}\ket{\Psi}\,,
\end{equation}
is minimized with respect to all states $\ket{\Psi}$ of the form (\ref{eq:MCSCF}), i.e.~with NON pinned to some specific polytope facet $F_D$.
Since this involves simultaneous optimization of the expansion coefficients
$\{c_{\bd{i}}\}_{\bd{i}\in \mathcal{I}_D}$ and the orbitals $\ket{i}$ this defines
a specific multiconfigurational self-consistent field (MCSCF) optimization.

At least in principle, this ansatz requires to reinforce the self-consistency
conditions on the coefficients $\{c_{\bd{i}}\}_{\bd{i}\in \mathcal{I}_D}$ according to
Remark \ref{rem:selfconsistent}. Yet, skipping them simplifies the variational optimization and can only lead to
better variational results. The minimization of the energy (\ref{eq:MCSCFenergy}) for the specific
MCSCF ansatz (\ref{eq:MCSCF}) is subject to the normalization of the quantum state, $\|\Psi\|_2=1$ and the orthonormality
of the orbitals $\ket{i}$ employed in (\ref{eq:MCSCF}). This, at least in principle, leads to several Lagrange multipliers.
However, in practice it has been proven advantageous to dodge the use of such Lagrange multipliers. We follow in \cite{CSHF}
and for the results shown in the next section those standard procedures from quantum chemistry \cite{pinkbook} which we recall briefly.

First, an orthonormal reference basis $\{\ket{\overline{i}}\}_{i=1}^d$ for the $1$-particle Hilbert space is fixed \footnote{In quantum chemistry, these are typically atomic orbitals and the Gram-Schmidt procedure may be used to orthonormalize them.}. This could be given by the Hartree-Fock molecular orbitals, if a Hartree-Fock optimization was preceding, or just the $1$-particle eigenstates of the external potential. Then, by expressing the orbitals $\ket{i}$ for ansatz (\ref{eq:MCSCF}) according to
\begin{equation}
\ket{i} = e^{\hat{\eta}} \ket{\overline{i}}\,,\quad \forall i\,,
\end{equation}
with an antihermitian operator $\hat{\eta}$, the orbital optimization is realized in form of an optimization of $\hat{\eta}$ and the unitarity of $e^{\hat{\eta}}$ makes the use of Lagrange multipliers obsolete.

Implementing this in (\ref{eq:MCSCFenergy}), using second quantization with respect to the fixed basis states $\{\ket{\overline{i}}\}_{i=1}^d$ and
\begin{equation}
\hat{\eta} \equiv \sum_{i,j=1}^d \eta_{ij} a_i^\dagger a_j, \qquad  \eta\equiv (\eta_{ij}),
\end{equation}
leads to the following form for the energy functional (\ref{eq:MCSCFenergy})
\begin{equation}
E[\{c_{\bd{i}}\}_{\bd{i}\in \mathcal{I}_D},\eta] = \bra{\overline{\Psi}}e^{-\hat{\eta}}\hat{H} e^{\hat{\eta}}\ket{\overline{\Psi}}\,,
\end{equation}
where
\begin{equation}
\ket{\overline{\Psi}} = \sum_{\bd{i}\in \mathcal{I}_D} c_{\bd{i}} \ket{\overline{\bd{i}}}
\end{equation}
and $\overline{\bd{i}}\equiv (\overline{i_1},\ldots,\overline{i_N})$. The variational optimization
with respect to $\eta$ leads to coupled generalized Hartree-Fock equations and the optimization of
the coefficients to so-called secular equations \cite{BenavThesis}. These self-consistent field equations can
iteratively be solved, e.g., by using the well-known Newton-Raphson optimization method,
one of the standard methods in quantum chemistry for solving MCSCF equations \cite{Shepard}.

In Ref.~\cite{CSHF}, we implement such a MCSCF algorithm and test it for a quantum dot system containing $N=3,4,5$ spin-polarized electrons. For the case $N=3$, remarkable $99.943\%$ of the correlation energy is attained and for $N=4,5$ about $99.941\%$ and $99.934\%$, respectively.
Since those few-fermion systems are known for their excellent quasipinning \cite{CSQ, CSthesis,CS2013,Ebler}
these results confirm our expectation that the MCSCF ansatz based on pinning works very well whenever the exact ground state exhibits at least quasipinning.

\section{Application to the lithium atom}\label{sec:Li}

In this section we apply the MCSCF ansatz based on the simplified pinning structure (\ref{eq:MCSCF}) to the lithium atom. As it is explained in Ref.~\cite{CSHF} there is a
hierarchy of settings $(3,d)$ that one can use for this. Here, we will use the smallest nontrivial setting $(N,d)=(3,6)$ and there is also no ambiguity for the choice of the facet since its polytope has only one nontrivial facet.

Furthermore, it should be stressed that in a spin-compensated configuration state basis set 
the spin-orbitals appear in pairs which may share the same spatial orbital
\footnote{A spin-compensated configuration state basis set is a basis set where the number
of spin orbitals with spin up equals the number of spin orbitals with spin down.}. Then, as it was shown in Ref.~\cite{BenavQuasi2} for the Borland-Dennis setting this implies pinning for GPC (\ref{eq:gpc36b}) in case of not too strong correlations.

Regardless of the spin symmetry, any wave function given as
superposition of Slater determinants \eqref{eq:BDstate}
tends to artificially mix different electronic spin-states.
Consequently, it is not an eigenfunction of the total spin-squared operator.
However, as long as the Hamiltonian commutes with the spin operator,
 solutions of the Schr\"odinger equation must be eigenfunctions of both
 spin operators $\mathbf{S}^2$ and $\mathbf{S}_z$. Since in general Slater
 determinants are not eigenfunctions of the total spin operator,
spin-adapted linear combinations of these determinants are usually imposed
to circumvent the problem of spin contamination \cite{Pauncz}. To ensure
a correct spin description of the state~\eqref{eq:BDstate} there are two
possible ways of assigning spatial orbitals $\phi_i$, $\bra{\phi_i}\phi_j\rangle = \delta_{ij}$,
\begin{align*}
\ket{1} &\equiv \phi_1(\rr)\up, \quad
\ket{2} \equiv \phi_2(\rr)\up, \quad
\ket{4} \equiv \phi_3(\rr)\up, \\
\ket{3} &\equiv \phi_1(\rr)\dn, \quad
\ket{6} \equiv \phi_2(\rr)\dn, \quad
\ket{5} \equiv \phi_3(\rr)\dn,
\end{align*}
or
\begin{align*}
\ket{2} &\equiv \phi_1(\rr)\up, \quad
\ket{1} \equiv \phi_2(\rr)\up, \quad
\ket{4} \equiv \phi_3(\rr)\up, \\
\ket{3} &\equiv \phi_1(\rr)\dn, \quad
\ket{5} \equiv \phi_2(\rr)\dn, \quad
\ket{6} \equiv \phi_3(\rr)\dn\, .
\end{align*}
The chosen configuration is the one which gives the lowest energy.

Recently rank-six, -seven and -eight approximations were accomplished
for the lithium isoelectronic series, by using a set
of helium-like $1$-particle wave functions and one hydrogen-like wave
function  \cite{BenavLiQuasi}. Shull and L\"owdin
employed for the former the following set of orthonormal spatial orbitals:
\begin{align} \label{eq:SHbasisset}
\delta_n(a, \rr) := D_n\sqrt{\frac{a^3}{\pi}}L^2_{n-1}\bigl(2 a
r\bigr)e^{-a r},
\end{align}
where $D^{-2}_n=\binom{n-1}{2}$ and $L^\zeta_n$ stands for the associated
Laguerre polynomials \cite{reliablerussian}, obtaining more than $50\%$ of
the correlation energy for the helium atom with only three spatial orbitals in the expansion~\cite{SL2}.
For the hydrogen-like function, the radial orbital
\begin{align} \label{eq:Hbasiset}
\chi_1(b,\rr) = \frac14\sqrt{\frac{b^5}{6\pi}}r \,
e^{-b r/2}
\end{align}
was used.

A standard variational procedure for the single determi\-nan\-tal state
$$
\ket{\delta_1 \up, \delta_1 \dn, \chi_1 \up}
$$ leads to the minimum for the energy at
$(a_0, b_0) = (2.6864, 1.2751)$.
The total energy associated to this Slater determinant
becomes $-7.4179$ a.u.~\cite{BenavLiQuasi}.
This is reasonably close to the Hartree-Fock energy $-7.4324$ a.u.~(computed
by using a cc-pVDZ basis set) or even better than the energy $-7.3815$
a.u.~(with a 3-21G* basis set) \cite{data}.  A highly accurate value for the
nonrelativistic ground-state energy of the lithium atom is $-7.478$
a.u.~\cite{ExactLitio}.

In the following, we minimize the energy by using the specific MCSCF ansatz given by
\eqref{eq:BDstate}. In particular this means to optimize
the natural spin orbitals $\{\ket{i}\}_{1\leq i \leq 6}$. The corresponding
spatial orbitals $\{\ket{\phi_i}\}_{1\leq i \leq 3}$
are expanded on a given finite basis. In this case, we choose
\begin{equation}
\phi_i(\rr) = \sum^M_{n=1} C_{ni} \delta_n(a_0,\rr)
+ \sum^2_{n=1} D_{ni} \chi_n(b_0,\rr)\,,
\label{eq:chiNO}
\end{equation}
in a similar fashion as in on-the-shelf SCF methods.
In order to obtain more correlation energy we have
added a second hydrogen-like radial wave function,
$$
\chi_2(b,\rr) = \frac8{81}\sqrt{\frac{b^7}{30\pi}}r^2 \,
e^{-b r/3}.
$$

Before implementing the full MCSCF optimization process, we
accomplished a Hartree-Fock pre-optimiza\-tion of the natural spin orbitals.
Pre-optimizations of this sort are not required in general
but are advised to increase the rate of convergence.
Once this initial step is completed,
the initial pinned-state ansatz \eqref{eq:MCSCF} can be constructed,
and the full \mbox{MCSCF} process can be implemented.
For $M =8$, we have obtained the energy $-7.472$ a.u.. This represents $87.09\%$ of the correct correlation of the lithium atom, which can be even improved by choosing larger $M$.

\section{Relation of quasipinning and reconstructed correlation energy}\label{sec:QPCorrE}
\begin{figure}[t]
\centering
\includegraphics[width=7.5cm]{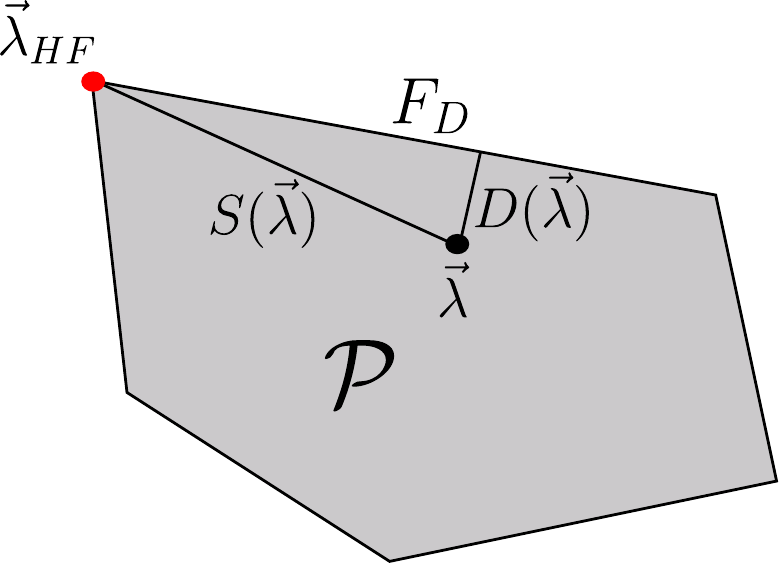}
\caption{Schematic illustration of the polytope $\mathcal{P}$
of possible vectors $\vec{\lambda}=(\lambda_i)$ of natural occupation numbers defined by the family of generalized Pauli constraints. For a given
$N$-fermion state we show the corresponding vector $\vec{\lambda}$ and
illustrate its minimal distance $D(\vec{\lambda})$ to the polytope facet $F_D$, as well as the distance $S(\vec{\lambda})$ to the Hartree-Fock point $\vec{\lambda}_{\rm HF}$ (\ref{eq:HF}).}
\label{fig:polygeom}
\end{figure}

It is expected that the variational ground state ansatz (\ref{eq:MCSCF}) based on pinning to the polytope facet $F_D$ leads to good results whenever the exact ground state has NON $\vec{\lambda}$ close to $F_D$. In this section we confirm this in form of rigorous mathematical estimates.

First we introduce a measure for the correlation of a quantum state $\ket{\Psi}$, given by the $l^1$-distance of $\vec{\lambda}$ to the Hartree-Fock point (\ref{eq:HF}),
\begin{equation}\label{eq:distHF}
S(\vec{\lambda}) \equiv \mbox{dist}_{l^1}(\vec{\lambda},\vec{\lambda}_{\rm HF}) =\sum_{i=1}^N(1-\lambda_i)+\sum_{i=N+1}^d \lambda_i\,.
\end{equation}

In the same way, a natural measure for quasipinning by a GPC $D$ was introduced in Ref.~\cite{CS2013}, namely the $l^1$-distance of $\vec{\lambda}$ to the corresponding polytope facet $F_D$ defined by pinning of $D$,
\begin{equation}
F_D \equiv \{\vec{\lambda}\in \mathcal{P}\,\mid\,D(\vec{\lambda})=0\}\,.
\end{equation}
As explained in \cite{CSQ} this distance coincides up to a factor $2$
with $D(\vec{\lambda})$ \footnote{The coefficients $\kappa^{(i)}, i=0,1,\ldots,d$ are always chosen as minimal integers. This fixes the ambiguity $D(\cdot)\geq 0 \Leftrightarrow \tilde{D}(\cdot)\equiv \alpha D(\cdot)\geq 0$,  $\alpha \in \R^+$.}. For an illustration see Figure \ref{fig:polygeom}.

%

\begin{figure}[t]
\centering
\includegraphics[width=3.7cm]{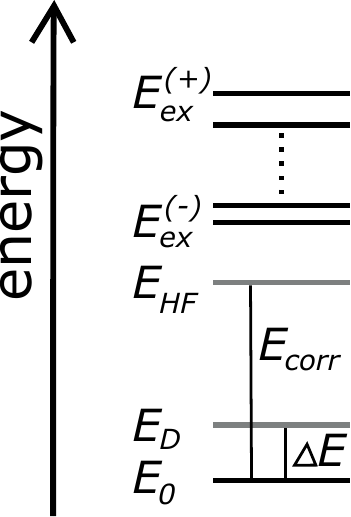}
\caption{Energy spectrum of the Hamiltonian $\hat{H}$ is shown in black. The ground state energy $E_0$ is assumed to be non-degenerate. All excitations of $\hat{H}$ lie in the interval $[E_{\rm ex}^{(-)},E_{\rm ex}^{(+)}]$. Also the Hartree-Fock ground state energy $E_{\rm HF}$ and the energy $E_D$ obtained by a MCSCF ansatz based on pinning to a facet $F_D$ are shown (in gray). Whenever the exact ground state exhibits quasipinning to $F_D$ we find $\Delta E\equiv E_D-E_0\ll E_{\rm HF}-E_0 = E_{\rm corr}$.}
\label{fig:energyspec}
\end{figure}

Let us now consider a Hamiltonian $\hat{H}$ acting on the $N$-fermion Hilbert space $\mathcal{H}=\wedge^N[\mathcal{H}_1^{(d)}]$. We assume that its ground state $|\Psi_0\rangle$ is unique and has the energy $E_0$. Since the Hilbert space is finite the ground state is therefore gapped and the set of all eigenenergies is bounded, i.e.~there exists a lowest excited energy $E_{\rm ex}^{(-)}> E_0$ and a maximal finite excitation energy $E_{\rm ex}^{(+)}$. The spectrum of the Hamiltonian is illustrated in Figure \ref{fig:energyspec}.

The numerical quality of our MCSCF ansatz based on the pinning structure (\ref{eq:MCSCF}) can then be estimated in the following way.
\begin{thm}\label{thm:correlationE}
Let $\hat{H}$ be a Hamiltonian on $\wedge^N[\mathcal{H}_1^{(d)}]$ with a unique ground state with NON $\vec{\lambda}=(\lambda_i)_{i=1}^d$ and energy spectrum as described in Figure \ref{fig:energyspec}. The error $\Delta E$ in the energy of the MCSCF ansatz based on pinning to a given facet $F_D$ of the polytope $\mathcal{P}$ is bounded from above,
\begin{equation}\label{eq:estimate1}
 \Delta E \leq C\, D(\vec{\lambda})\,,
\end{equation}
with
$C=\tilde{C}\,(E_{\rm ex}^{+}-E_0)$ for some constant $\tilde{C}$.
Moreover,
\begin{equation}\label{eq:estimate2}
\frac{\Delta E}{E_{\rm corr}} \leq K\, \frac{D(\vec{\lambda})}{S(\vec{\lambda})}\,,
\end{equation}
where $E_{\rm corr}\equiv E_{\rm HF}-E_0$ is the correlation energy and $K=\frac{E_{\rm ex}^{(+)}-E_0}{E_{\rm ex}^{(-)}-E_0}\,\frac{\tilde{C}}{N}$.
\end{thm}
\noindent The proof of the above Theorem
and further details can be found in Ref.~\cite{CSHF}.
Estimate \eqref{eq:estimate1} confirms our expectation that the
MCSCF ansatz based on pinning works very well whenever the exact
ground state exhibits quasipinning. Intriguingly, this can be geometrically
quantified by referring to the distance of the exact ground state NON
$\vec{\lambda}$ to the facet $F_D$. Another important estimate on the
numerical quality is provided by estimate \eqref{eq:estimate2}. It states that
the fraction of the correlation energy which is \emph{not} covered by the MCSCF
ansatz, is bounded from above by the ratio $D(\vec{\lambda})/S(\vec{\lambda})$.
Hence, the overwhelming part of the correlation energy is reconstructed
whenever the quasipinning is nontrivial in the sense $D(\vec{\lambda}) \ll S(\vec{\lambda})$.

\section{Summary and conclusion}\label{sec:concl}

In this paper we have discussed the recently proposed
natural extension of the Hartree-Fock ansatz based on
the remarkable structural simplifications given saturation of
some generalized Pauli constraints. This ansatz defined through extremal
$1$-fermion information is actually a method for selecting non-superfluous
configurations in Configuration Interaction computations.
This MCSCF variational ansatz has already been tested for a
quantum dot system containing three, four or five spin-polarized
electrons, reaching  more than $99.9\%$ of the correlation
energy~\cite{CSHF}. In this paper we have used the Borland-Dennis
ansatz for computing the ground-state energy of the lithium atoms
by means of a very economic wave functions built from only three
Slater determinants.

We have also discussed in some detail the relation between
the $N$-fermion wave function and its reduced $1$-fermion description.
In particular we have studied the underlying mathematical structure of
quasipinning and provided a universal geometrical bound on the
correlation energy reconstructed by the corresponding pinning
ansatzes. Forthcoming work is devoted to the derivation of
universal geometric bounds on
the numerical quality of commonly used methods in quantum
chemistry as, e.g.,
Hartree-Fock and CASSCF. By employing gradient methods, we also gain further insights on the
preimage of moment maps and the implications of local information
extremal within the moment polytope \cite{CSHF}.

\section*{Acknowledgements}
We are grateful to Jos\'e M.~Gracia-Bond\'ia,
Dieter Jaksch, Nektarios Lathiotakis, Miguel Marques and Vlatko Vedral
for helpful discussions. Financial supported is acknowledged from the Colombian
Department of Sciences and Technology (CBR) and from the Swiss National Science Foundation (Grant P2EZP2 152190) and the Oxford Martin Programme on Bio-Inspired Quantum Technologies (CS).

\bibliography{Birthday_revision}

\end{document}